\documentclass[lettersize,journal]{IEEEtran}
\usepackage{amsmath,amsfonts}
\usepackage{algorithmic}
\usepackage{algorithm}
\usepackage{array}
\usepackage[caption=false,font=normalsize,labelfont=sf,textfont=sf]{subfig}
\usepackage{textcomp}
\usepackage{stfloats}
\usepackage{url}
\usepackage{verbatim}
\usepackage{graphicx}
\usepackage{cite}
\usepackage{multirow}
\usepackage{array} 
\begin{document}

\title{Compact Plasmonic Logic Gates Enabled by Magnetoelectric Light Funneling for On-Chip Optical Computing in the Telecom Band}

\author{Adib Md. Tawsif$^{1,\&}$, Khondokar Zahin$^{1,\&}$, A.K.M. Hasibul Hoque$^{1}$, Ying Yin Tsui$^{2}$, and Md Zahurul Islam$^{1,*}$
\thanks{$^{1}$ Department of Electrical and Electronic Engineering, Bangladesh University of Engineering and Technology, ECE Building, West Palashi Campus, Dhaka 1205, Bangladesh}
\thanks{$^{2}$ Department of Electrical and Computer Engineering, University of Alberta, Edmonton, AB T6G 2H5 Canada}
\thanks{$^{\&}$ Authors with equal contributions}
\thanks{$^{*}$ Corresponding Author, email: mdzahurulislam@eee.buet.ac.bd}
}



\maketitle

\begin{abstract}
The realization of all-optical logic gates (AOLGs) is important for advancing photonic integrated circuit (PIC) design and optical data communication. Various photonic structures and design techniques, including two-dimensional photonic crystals, silicon waveguides, plasmonic waveguides, diffractive neural networks, and inverse design techniques, are actively being explored to achieve multifunctional, high-performance optical logic gates with fast data processing capabilities. Among these, plasmonic structure-based AOLGs often face challenges such as fabrication complexity and large device footprints when integrating multiple logic operations within a single structure. In this work, a planar and compact metal-insulator-metal (MIM) plasmonic waveguide structure is proposed for AOLG design, utilizing the light funneling effect in grooved metasurfaces. The designed device, with dimensions of 700 nm × 460 nm, successfully implements three fundamental logic gates (NOT, AND, OR) with a high contrast ratio of 18.69 dB. The device operates within the wavelength range of 1400 nm to 1450 nm, making it suitable for the telecommunications field. Its planar architecture offers fabrication feasibility and all logic gates can be controlled using a single light source incident from one specific direction, which facilitates its integration into photonic circuits with fast operational speed. This work contributes to the advancement of scalable and high-speed photonic computing platforms. 
\end{abstract}

\section{Introduction}
Photonic Integrated Circuits (PICs) have come a long way in overcoming the challenges that electronic integrated circuits (EICs) face, due to their low power consumption and high-speed operation\cite{ref1}. Optical logic devices serve as the building blocks of photonic integrated circuits (PICs) and optical communication. To analyze all optical logic gates (AOLG), different types of photonic structures have been proposed. For example, reconfigurable optical logic gates (AND, NOR, NOT) have been constructed in 2D photonic crystal square lattice, enabling various optical logic functionalities within a single platform\cite{ref2,ref3,ref4}. Diffractive optical neural networks (DONNs) and topologically protected optical and quantum logic gates are being proposed for the robust device structure\cite{ref5,ref6,ref7,ref8}. Besides, semiconductor optical amplifiers, cavities of a hybrid nano-crystalline beam, two-photon absorbers in silicon waveguides, ring resonators, and phase modulators have been designed for the realization of AOLGs\cite{ref9,ref10,ref11,ref12,ref13}. Recently, nanoscale optical logic gates based on plasmonic structures have drawn attention because of their ability to process light at the sub-wavelength scale\cite{ref14}.  For achieving compact footprints in metal-insulator-metal (MIM) waveguide-based plasmonic logic gates, multi-objective optimization using the inverse design method has been reported\cite{ref15}. They have reported AOLGs in a single structure of 0.8 \(\mu\)m × 1.1 \(\mu\)m with different extinction ratios for different gates operating at 1.31 \(\mu\)m wavelength. Haffer et al. reported a MIM waveguide coupled with an elliptical ring resonator for AOLGs, achieving a high contrast ratio\cite{ref16}. Another ring resonator-based plasmonic nanostructure design for all-optical NOR and NOT gates has been reported\cite{ref17} high transmission based on the shape of the ring resonators. Optical logic NOT gate based on a plasmonic MIM waveguide structure of 2.4 \(\mu\)m × 3 \(\mu\)m dimension was numerically demonstrated, achieving a maximum transmission of 65.35\% by controlling the propagation through the control port\cite{ref18}. Two plasmonic waveguide structures incorporating MIM square ring resonators were proposed to implement NOT, AND, and NOR gates, with compact footprints of 750 nm × 900 nm (NOT) and 1.5 \(\mu\)m × 1.8 \(\mu\)m (AND/NOR) at 1535 nm, respectively\cite{ref19}. Feynman logic gate, based on cascaded MIM optical waveguides with Mach–Zehnder interferometers, achieves an extinction ratio of 10.57 dB and low insertion loss, within a compact footprint of a few square micrometers\cite{ref20}. Most of the above-mentioned studies have attempted to minimize the footprint by integrating a small number of logic gates into a single structure or altering some dimensions of the device to increase the number of logic gates. Additionally, these kinds of designs have fabrication complexity, and multiple light sources from different directions need to be used for different slots for the interference of light trapped in the dielectric layer. Our study proposes three basic plasmonic logic gates—NOT, AND, and OR—based on MIM waveguides utilizing the magnetoelectric light funneling phenomenon within a single grooved metasurface-based nanostructure. The proposed design features a compact footprint and supports flexible planar fabrication techniques. It operates at a single wavelength in the optical communication bandwidth, which is tunable by adjusting device parameters. The device operates using a single input light source from a single direction at multiple grooves, significantly reducing the footprint for photonic circuit integration while improving optical data processing speed and energy efficiency.
\newpage
\section{Light Funneling in Grooved Meta Surface}
\begin{figure}[h]
    \centering
    \includegraphics[width=1\linewidth]{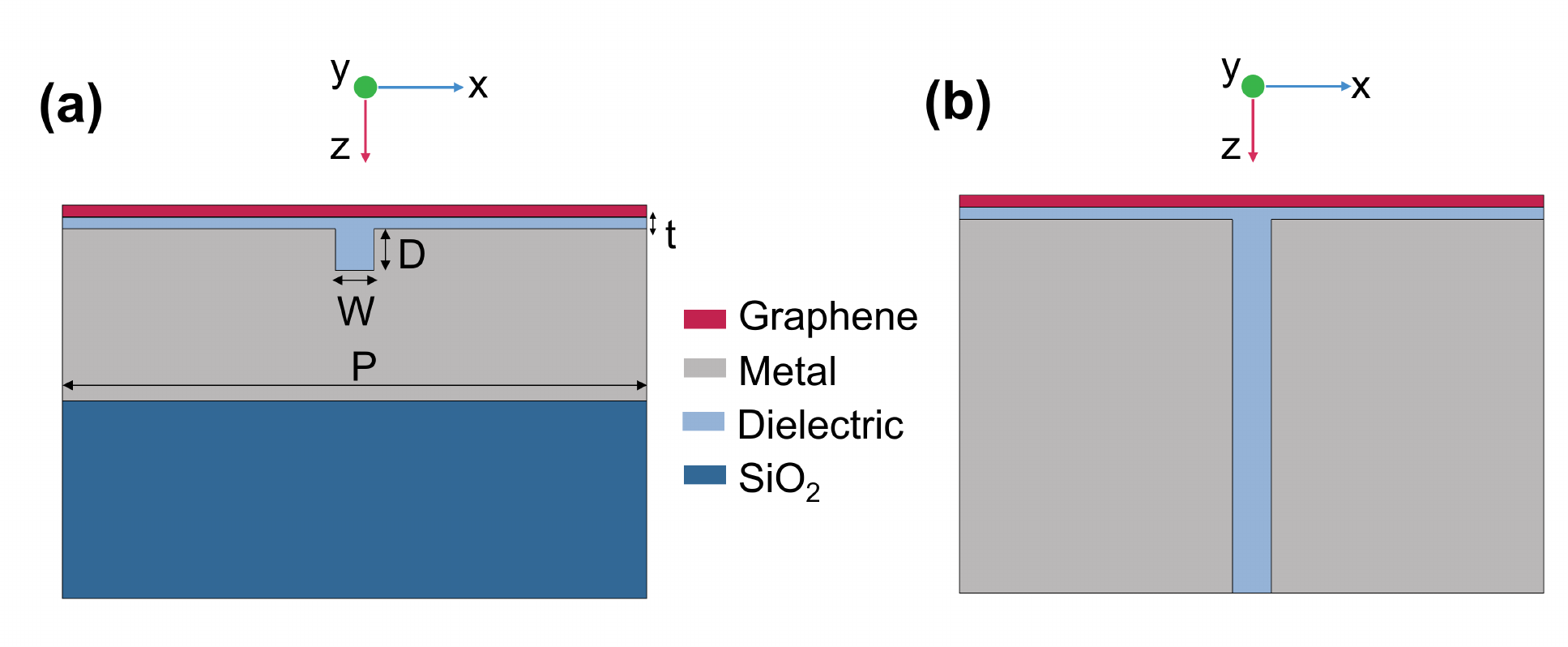}
    \caption{(a) Light funneling inside the grooved metasurface consisting of \(Ag\) as metal layer, dielectric groove of width W and depth D and graphene layer above to increase absorbance. (b) To funnel light out of the groove, it is made open-ended by removing the substrate layer which increase transmittance.}
    \label{fig:1}
\end{figure}

Pablo et al. first reported the study and design of nanophotonic devices relying on the light funneling phenomenon\cite{ref21}. According to the study, light funneling results from magnetoelectric interference(MEI) of the incident and the evanescent fields in both resonant and non-resonant conditions. In Fig.~\ref{fig:1}, a single groove of depth \(D\) and width \(W\) is etched on a metal film. Then a dielectric is deposited in the groove. The groove acts as a Fabry-Perot (FP) resonator that traps optical energy of specific wavelengths.
According to the theory of light funneling, light should funnel into the metallic grooves and get trapped in the FP resonant cavity if it maintains the following equation,
\begin{equation} \label{eq:1}
\left(\frac{1}{4}+\frac{1}{2} m\right) \lambda=n_{e f f} D
\end{equation}
where \(m\) is a positive integer and \(n_{eff}\) is the effective refractive index of MIM waveguide modes. The effective index is insignificantly dependent on the depth of the nanogroove \(D\), but is highly dependent on acute changes in width \(W\)\cite{ref22,ref23}.  To incorporate graphene with light funneling, multiple layers of graphene were stacked on the surface of the metallic groove for strong absorption in the mid-infrared region. Both studies of light funneling in and out have been carried out to optimize the design of optical logic gates, as shown in the structures of Fig.~\ref{fig:1}(a-b). Primarily, Silver's\((Ag)\) Palik model\cite{ref24} is used as a metal, and a polymethyl methacrylate (PMMA) spacer is used as a dielectric, which helps efficient light coupling. The refractive index of PMMA is taken as 1.49\cite{ref25}. The thickness of \(Ag\) is kept at 50 nm under the groove to obstruct the transmission of light through the groove completely. There is a sheet of PMMA of thickness t on the top of the groove, above which the layers of graphene are stacked. The whole structure is based on a \(SiO_2\) substrate. 
\begin{figure}[t]
    \centering
    \includegraphics[width=1\linewidth]{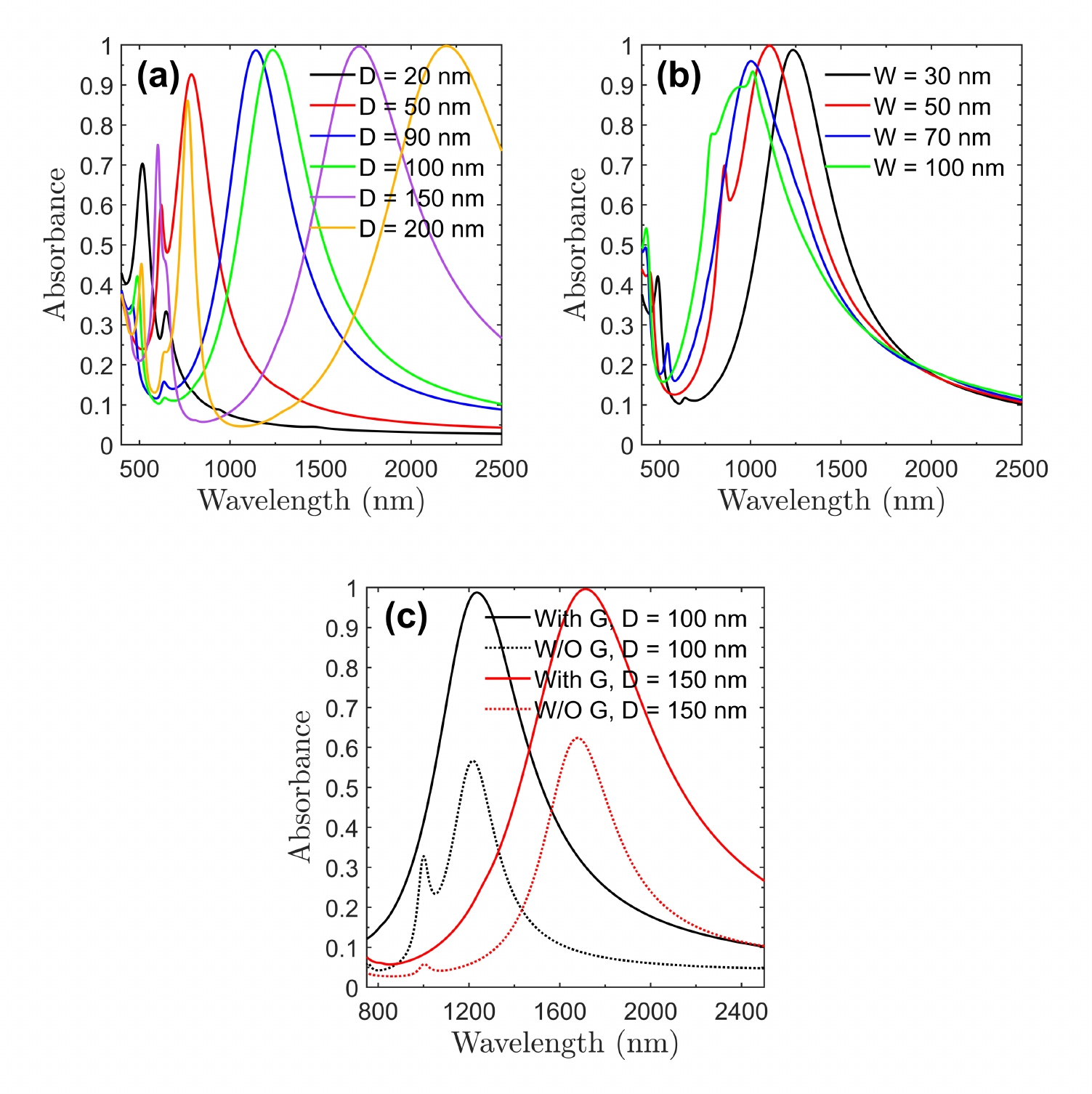}
    \caption{ Absorbance spectra of light which indicates the amount of light funneled inside: (a) For different groove depths of D = 20 nm, 50 nm, 90 nm, 100 nm, 150 nm, 200 nm (b) For different groove widths of W = 30 nm, 50 nm, 70 nm, 100 nm (c) With and without graphene as top layer for groove depth of 100 nm and 150 nm}
    \label{fig:2}
\end{figure}
A transverse magnetic (TM) polarized plane wave broadband source of 450 nm - 2500 nm was used for the study. The Finite Difference Time Domain (FDTD) method was employed to solve vector 3D Maxwell equations for studying the light funneling characteristics of a single-grooved metasurface. As we have used a metal film of thickness 50 nm at the bottom, zero transmittance is achieved from the groove. As shown in Fig.~\ref{fig:2}(a-b), when light is funneled into the groove, the absorption spectrum, which is an indicator of how much light is trapped or funneled inside the groove, exhibits a red shift with increasing groove depth and a blue shift due to increased groove width. It is also evident from Fig.~\ref{fig:2}(c) that removing the graphene layer from the top significantly reduces the amount of light energy trapped inside the groove. 
\begin{figure}[h]
    \centering
    \includegraphics[width=1\linewidth]{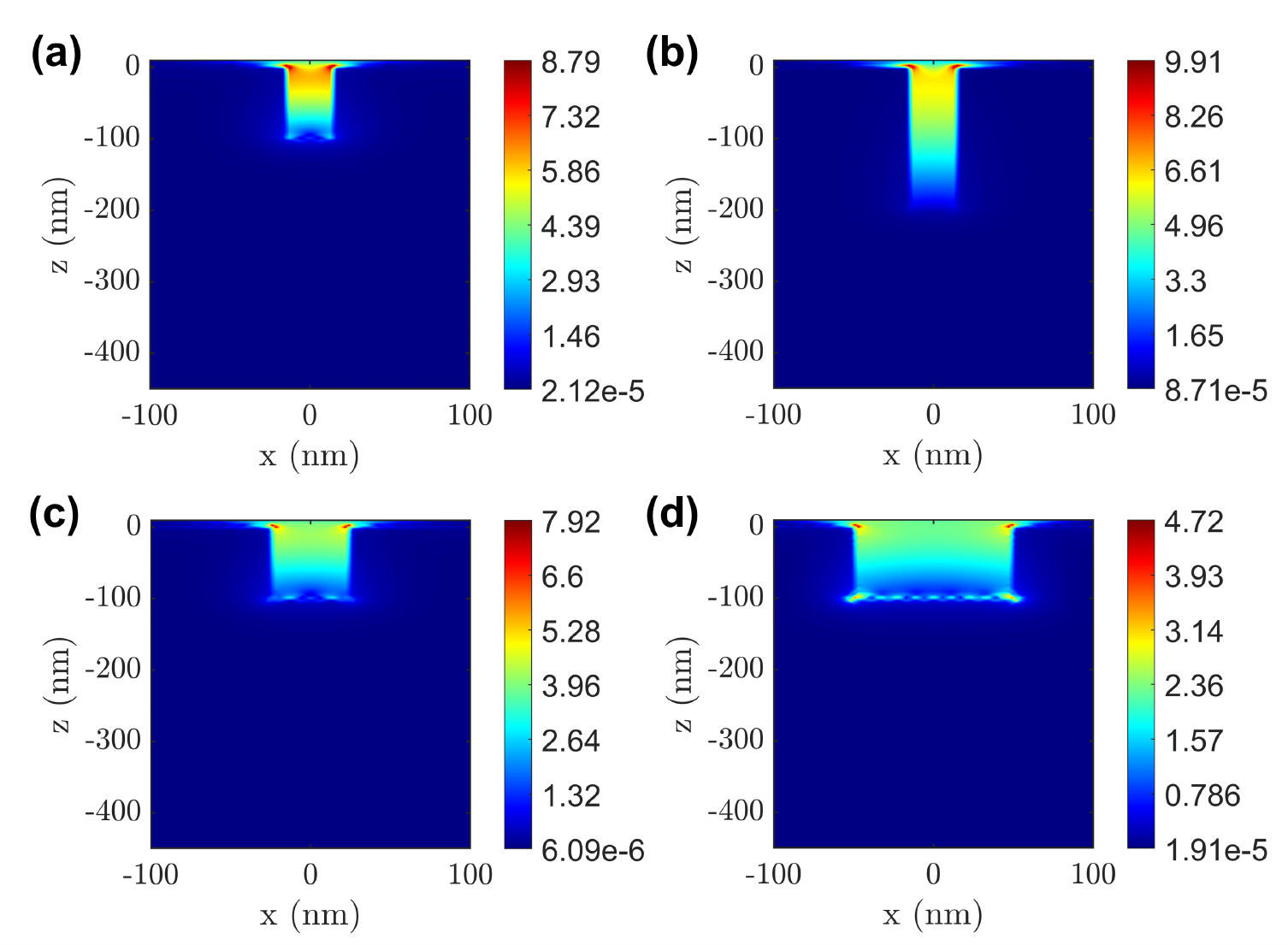}
    \caption{Variation in electric field intensity while the light funnels inside the groove with changing groove depth (a) D = 100 nm, (b) D = 200 nm and groove width (c) W = 50 nm, (d) W = 100 nm. Dipole resonance mode is visible at the corner in all cases alters the electric field direction into the groove and trap the light inside}
    \label{fig:3}
\end{figure}
It can be seen from Fig.~\ref{fig:3} that, around the corner of the dielectric groove, there is a significant concentration and enhancement of the electric field in all cases, which resembles an electric dipole resonance mode. The dipole alters the electric field of the incident light and further redirects the light into the groove. The electric field profiles in Fig.~\ref{fig:3}(a-d) show that the deeper grooves have stronger field magnitudes than the shallow grooves, and the wider grooves have weaker field magnitudes than the narrower grooves. 
\begin{figure}[h]
    \centering
    \includegraphics[width=1\linewidth]{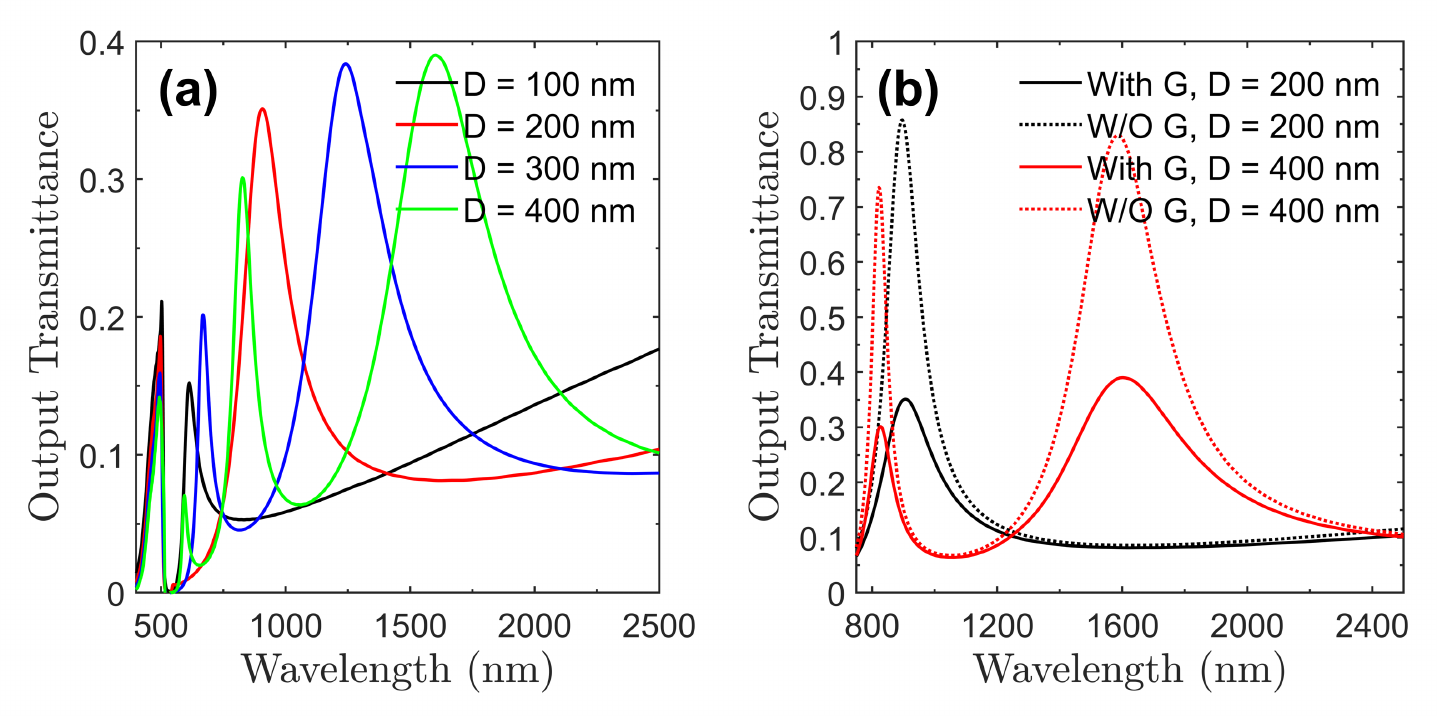}
    \caption{Transmission spectra of light that funneled out of the open-ended groove: (a) Varying groove depth of D = 100 nm, 200 nm, 300 nm, 400 nm (b) With and without graphene as top layer for groove depths of 200 nm and 400 nm. Graphene significantly reduces the transmission of funneled out light.}
    \label{fig:4}
\end{figure}
If the light that is funneled in could be funneled out through the other end, this phenomenon could be used in device-specific applications like optical logic gates. As a single mode of light is dominantly absorbed in the groove, it cannot readily be assured that a single mode of light will funnel out. Since the funneling out phenomenon is studied in an open-ended groove of reasonable width, a groove width of w = 100 nm is considered for this study by removing the bottom layer, as shown in Fig.~\ref{fig:1}(b). 
\begin{figure}[h]
    \centering
    \includegraphics[width=1\linewidth]{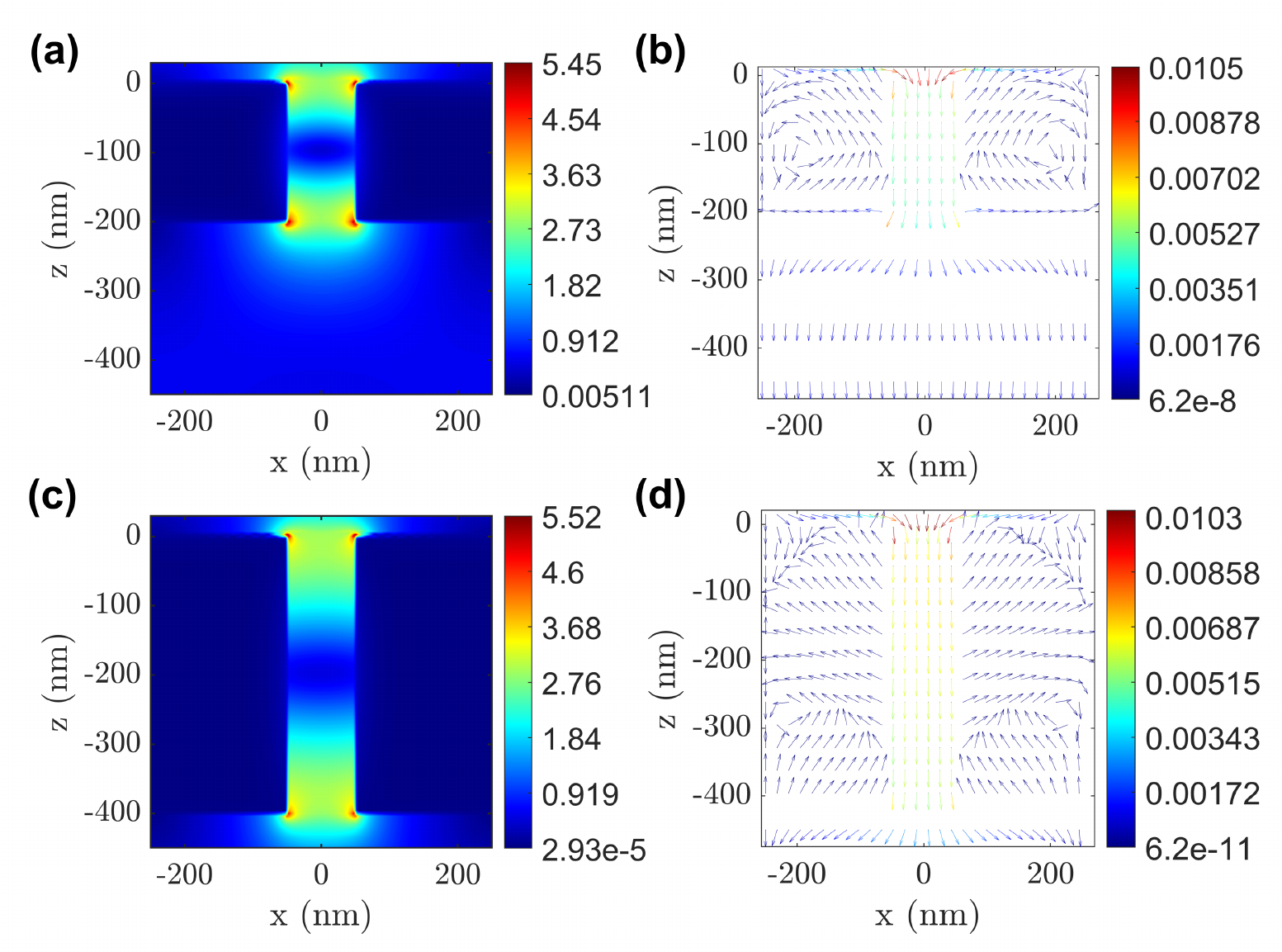}
    \caption{Variation in electric field intensity and optical power vector flow when light funnels out of the structure depends on the groove depth: (a-b) Groove depth of 200 nm (c-d) Groove Depth of 400 nm}
    \label{fig:5}
\end{figure}
In Fig.~\ref{fig:4}(a), it can be observed that the transmittance value increases with increasing groove depth because deeper grooves absorb more light, and thus they funnel out more compared to shallow grooves. The transmittance curve also becomes red-shifted with the increasing groove depth. Graphene significantly reduces the transmission of the funneled out light because of its absorption dominance in the wavelength range of the study, as shown in Fig.~\ref{fig:4}(b). 

To get a better understanding, the electric field profiles and their corresponding optical power vector flow (\(P = E X H\)) are observed in Fig.~\ref{fig:5}(a-d) for groove depths of 200 nm and 400 nm at the maximum transmission wavelength.

\section{Design of Plasmonic Logic Gates Leveraging Light Funneling Phenomenon}
\begin{figure}[h]
    \centering
    \includegraphics[width=1\linewidth]{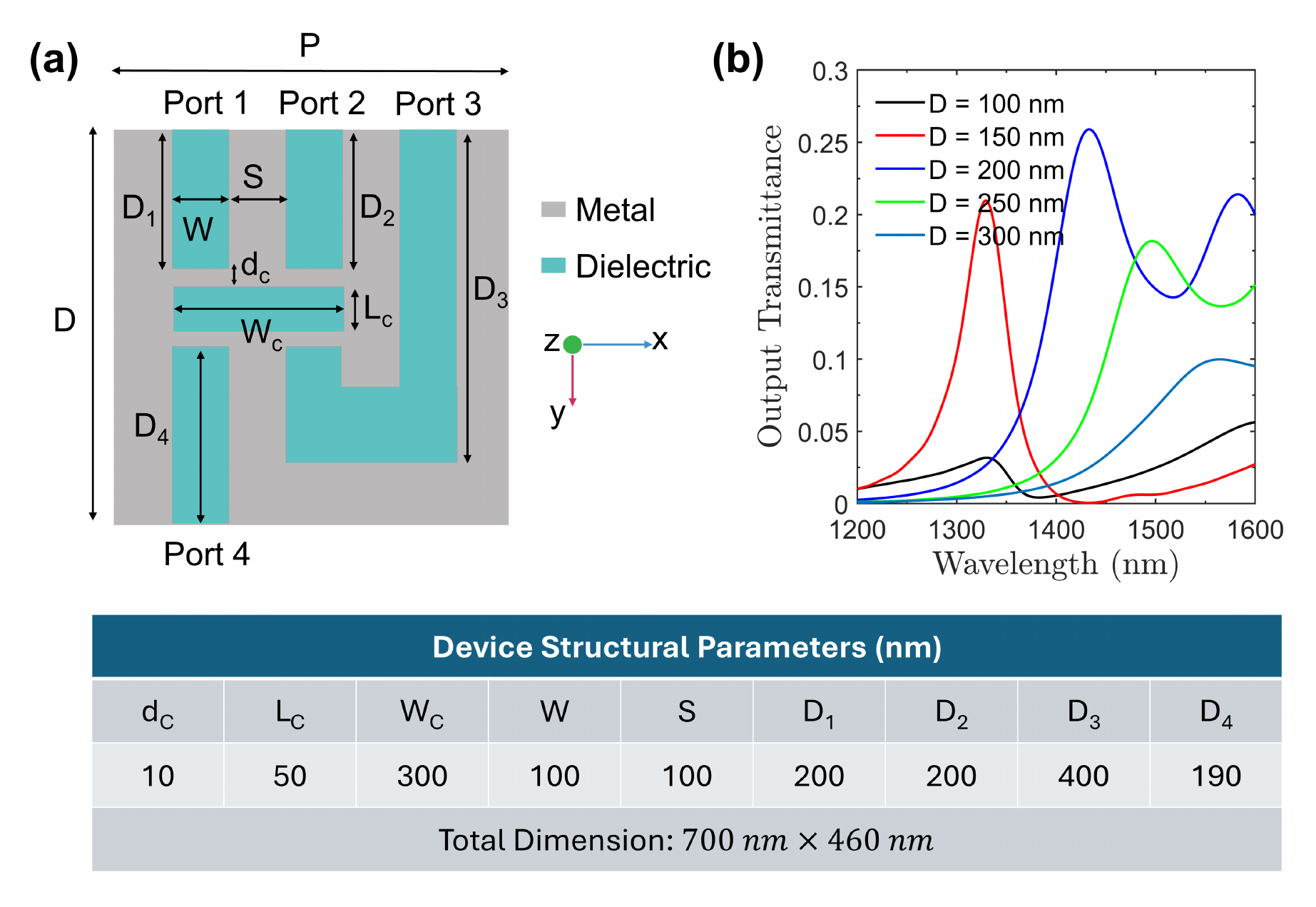}
    \caption{(a) Schematic of the top view of the proposed device with three dielectric grooves inside the metal for implementing fundamental logic gates (NOT, AND, OR) (b) Transmission spectra by varying groove depth of input port 1 (\(D_1\)), which is equal to the depth of port 2 ((\(D_2\)). Since the groove depth of 200 nm shows significant transmittance of light in the operating region, device parameters are fixed according to that. All the structural parameters of the proposed device are listed in the table}
    \label{fig:6}
\end{figure}

From Equation \ref{eq:1}, resonant modes of funneled light inside MIM waveguide depend on the effective refractive index of the waveguide,\(n_{eff}\). \(n_{eff}\) can be calculated using the formula based on the theory of plasmonics\cite{ref26},
\begin{equation}\label{eq:2}
\tanh \left(\frac{w \sqrt{\beta^2-k_o^2 \epsilon_d}}{2}\right)=-\frac{\epsilon_d \sqrt{\beta^2-k_o^2 \epsilon_m}}{\epsilon_m \sqrt{\beta^2-k_o^2 \epsilon_d}}
\end{equation}
where \(\epsilon_m\) and \(\epsilon_d\) are the dielectric constants of metal and the dielectric material, \(k_o\) is the wave vector of the incident light, and \(\beta\) is the propagation constant of light
through the MIM waveguide. \(n_{eff}\) can be calculated from here by the following relation,
\begin{equation}\label{eq:3}
n_{e f f}=\frac{\beta}{k_o}
\end{equation}
For designing three AOLGs (NOT, AND, OR) in a single structure, a three-grooved metasurface is proposed. The three grooves resemble three input ports which end up in a coupling slab, as shown in Fig.~\ref{fig:6}(a). 
The optical energy trapped inside the slab can be outcoupled through an output port. Coupling of light in the slot cavity takes place upon the fulfillment of the following condition\cite{ref27},
\begin{equation}\label{eq:4}
\Delta_\phi=2 \beta_m L_c+\phi_r=2 m \pi
\end{equation}
where \(\Delta\)\(\phi\) is the total phase displacement of the light for each transport round occurring in the slot cavity and \(\phi_r\) is the phase displacement on both sides of the cavity boundary when the wave propagates, \(L_c\) is the coupler slab’s length, m is the order of resonance in the slab, and \(\beta_m\) is the propagation constant corresponding to the \(m^th\) order mode. Thus, the resonant wavelength in the cavity can be expressed as\cite{ref27},
\begin{equation}\label{eq:5}
\lambda_m=\frac{2 n_{e f f} L_c}{m-\frac{\phi_r}{\pi}}
\end{equation}
For the simulation of the structure, the same materials are used for the metal and dielectric layers mentioned in the previous section. Since graphene and metal both materials show high absorption in the case of funneling the light inside, we removed graphene stacks from the top for logic gate design. A TM polarized wave of 1200 nm - 1600 nm wavelength is injected backward in the z-direction. Since the single light source is injected from one direction, it reduces the footprint of the integrated circuit and increases the speed of light transmission. Since metal structure significantly absorbs light, the depth of the grooves needs to be optimized for maximum transmittance. The device structure and parameter values are listed in Fig.~\ref{fig:6}. By varying \(D_1\) for maximum transmittance, the final device parameters are obtained. From Fig.~\ref{fig:6}(b), it is evident that only groove depths of 150 nm and 200 nm provide acceptable transmittance levels. To keep the operating wavelength from 1400 nm - 1450 nm range, the groove depths for ports 1 and 2 are kept at \(D_1\) = \(D_2\) = 200 nm and \(D_3\) = 400 nm. So, the device dimensions become: 700 nm \(X\) 460 nm.

\begin{figure*}[t]
    \centering
    \includegraphics[width=1\linewidth]{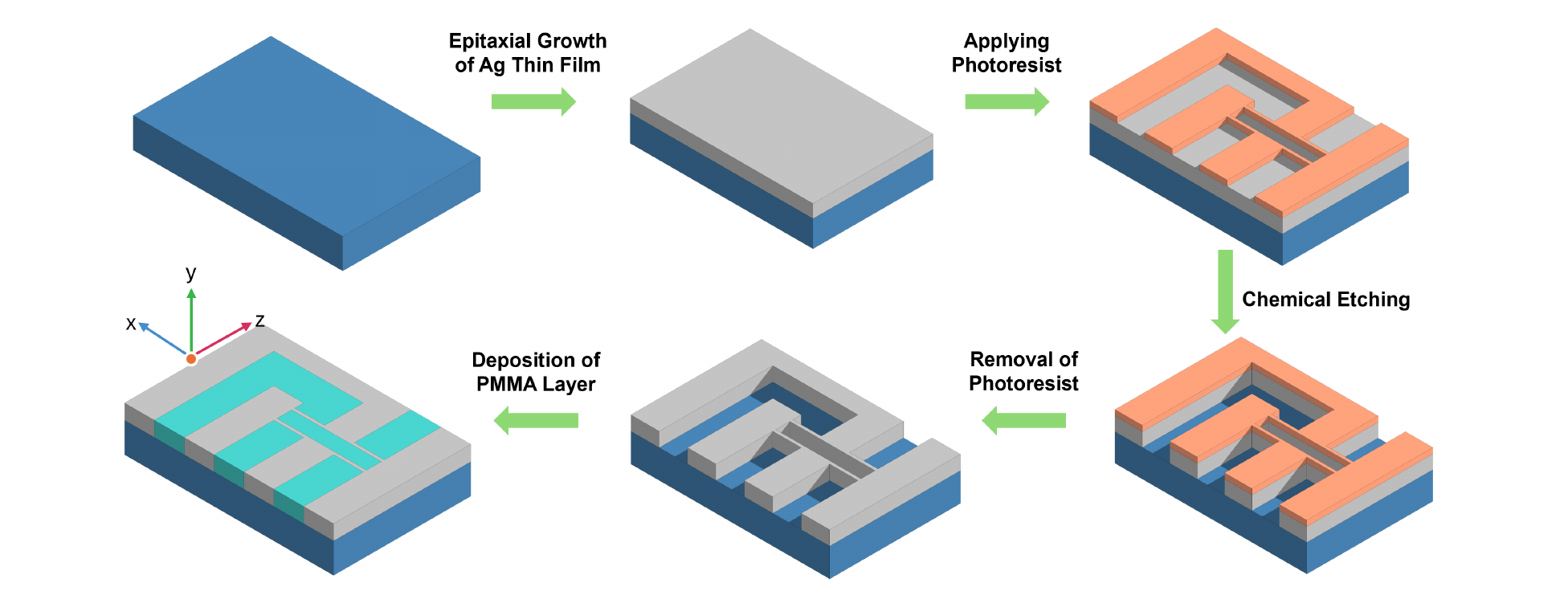}
    \caption{Device fabrication begins with the epitaxial deposition of a silver (\textit{Ag}) thin film on the substrate. Following this, a photoresist is applied to define the desired pattern, and the structure is etched using a selective chemical etchant. Finally, area-selective atomic layer deposition is employed to fill the patterned gaps with a PMMA layer.}
    \label{fig:com}
\end{figure*}
\section{Fabrication Process}
For fabricating the dielectric patterns on the metal epitaxial layer, techniques like photolithography and nanoimprinting can be used. Patterning of gold (\textit{Au}) with additive and subtractive methods has been discussed in the literature\cite{ref28,ref29,ref30}. In the subtractive method, dry or wet etching can be used where the metal film is covered with a lithographically patterned mask material and then exposed to a liquid or gas-phase etching agent. The wet etching method has the disadvantage of having the photoresist mask undercut since the etch proceeds isotropically. When the device size becomes 3 \(\mu m\) to 5 \(\mu m\), the feature loss issue becomes serious\cite{ref30,ref31}. Dry etching methods, such as reactive ion etching (RIE), have replaced wet etching for patterning on metal surfaces\cite {ref32,ref33}, although some metals, including gold, copper, and platinum, are challenging to dry etch. Another process used to pattern a metal surface is by applying mechanical pressure, which is also known as nanoimprinting. A substrate mold with photoresist creates pattern in the metal thin film with pressure applied from outside\cite{ref34}. For our device fabrication, the electron beam lithography or photolithography process is proposed here in Fig.~\ref{fig:com}.
Epitaxial growth of silver (\textit{Ag}) can be achieved through chemical vapor deposition (CVD) process or atomic layer deposition\cite{ref35,ref36,ref37}. Then, photoresists can be applied as a masking layer to develop the pattern on \textit{Ag} thin film. After exposure and development, the hardened mask defines the areas to remain. The metal (\textit{Ag}) lift off can be done by immersing the layers in an etchant solution. For etching of \textit{Ag}, etching solutions require a component that oxidises silver, and a further one that dissolves silver oxide. Silver can be etched using a solution composed of ammonium hydroxide (\(NH_4OH\)), hydrogen peroxide (\(H_2O_2\)), and methanol in a 1:1:4 volume ratio. Although methanol provides etching uniformity, water can be replaced because of the toxicity of methanol. Another viable etchant for silver is a 1:1:1 aqueous mixture of nitric acid (\(HNO_3\)), hydrochloric acid (HCl), and water\cite{ref38}. After detecting the endpoint of etching, the wafer is thoroughly rinsed (often with DI water) to stop the reaction, and the resist mask is stripped. After removing photoresist, to fill the gap with PMMA dielectric, the area selective atomic layer deposition (ALD) method can be used to fabricate the final device\cite{ref39,ref40}.

\section{Results and Analysis}
In this section, we will implement three basic gates with the mentioned device parameters and measure the operating wavelength and contrast ratio for each of the gates. Contrast ratio can be achieved with the following formula,
\begin{equation}\label{eq:5}
\mathrm{T}(\mathrm{~dB})=10 \log _{10}\left(\frac{P_{\mathrm{ON}}}{P_{\mathrm{OFF}}}\right)
\end{equation}
where, \(P_{ON}\) is the output optical power when the logic output is 1 and \(P_{OFF}\) is the output optical power when the logic output is 0

\subsection{NOT Gate}
For implementing the NOT gate, port 1 is used as input, port 2 is used as a control port, port 3 is unused and the output is observed from port 4. Proper functionality is achieved only when the control pin is high, which is listed in Table.~\ref{tab:1}.

\begin{table}[h!]
\centering
\caption{Truth table for implementing optical NOT gate}
\renewcommand{\arraystretch}{1.5}
\begin{tabular}{|c|c|c|c|>{\centering\arraybackslash}p{1.7cm}|}
\hline
\shortstack{\rule{0pt}{3ex}\textbf{Control}\\\textbf{Port}} & \shortstack{\textbf{Input}\\\textbf{(A)}} & \shortstack{\textbf{Output}\\\textbf{(B)}} &
\shortstack{\textbf{Contrast Ratio}\\\textbf{(T)}} \\
\hline
\multirow{2}{*}{High} & 0 & 1 & \multirow{2}{*}{13.8 dB} \\
                      & 1 & 0 & \\
\hline
\end{tabular}
\label{tab:1}
\end{table}

\begin{figure}[h]
    \centering
    \includegraphics[width=1\linewidth]{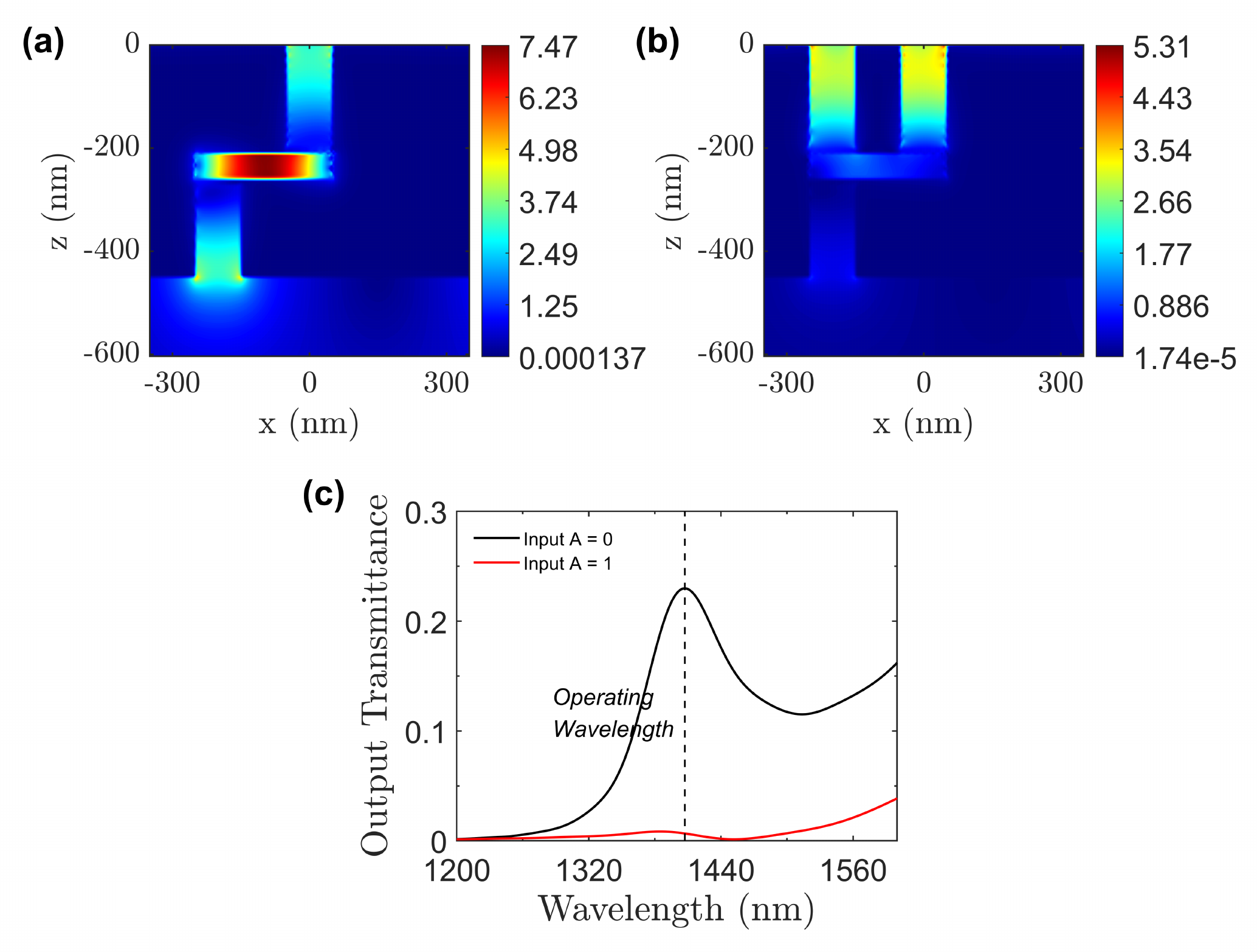}
    \caption{Electric field profile for NOT gate representation: (a) Input port 1 $->$ A=0, output=1 (b) Input port 1 $->$ A=1, output=0 (c) Output transmission spectrum for both inputs with the maximum value of 24\% at 'ON' state and the minimum of 1\% at 'OFF' state}
    \label{fig:7}
\end{figure}

From the transmittance curve in Fig.~\ref{fig:7}(d), it can be seen that the designed NOT gate is operational at \(\lambda\) = 1410 nm. For input, A = 0, 24\% transmittance is obtained at the output, whereas for A = 1, the transmittance is nearly 1\%. Contrast ratio for NOT gate is measured 13.8 \(dB\). The electric Field profile from Fig.~\ref{fig:7}(a-b) validates the functionality of the NOT gate.

\subsection{AND Gate}
For implementing the AND gate, port 1 is used for the first input(A), port 2 is used as the control port and port 3 is used for the second input(B). The output (C) is observed from port 4. Proper functionality is achieved only when the control pin is high, which is listed in Table.~\ref{tab:2}.
\begin{table}[h!]
\centering
\caption{Truth table for implementing optical AND gate}
\renewcommand{\arraystretch}{1.5}
\begin{tabular}{|c|c|c|c|>{\centering\arraybackslash}p{1.7cm}|}
\hline
\shortstack{\rule{0pt}{3ex}\textbf{Control}\\\textbf{Port}} & \shortstack{\textbf{Input}\\\textbf{(A)}} & \shortstack{\textbf{Input}\\\textbf{(B)}} & \shortstack{\textbf{Output}\\\textbf{(C)}} &
\shortstack{\textbf{Contrast Ratio}\\\textbf{(T)}} \\
\hline
\multirow{4}{*}{High} & 0 & 0 & 0 & \multirow{4}{*}{14.31 dB} \\
                      & 0 & 1 & 0 & \\
                      & 1 & 0 & 0 & \\
                      & 1 & 1 & 1 & \\
\hline
\end{tabular}
\label{tab:2}
\end{table}

\begin{figure}[h]
    \centering
    \includegraphics[width=1\linewidth]{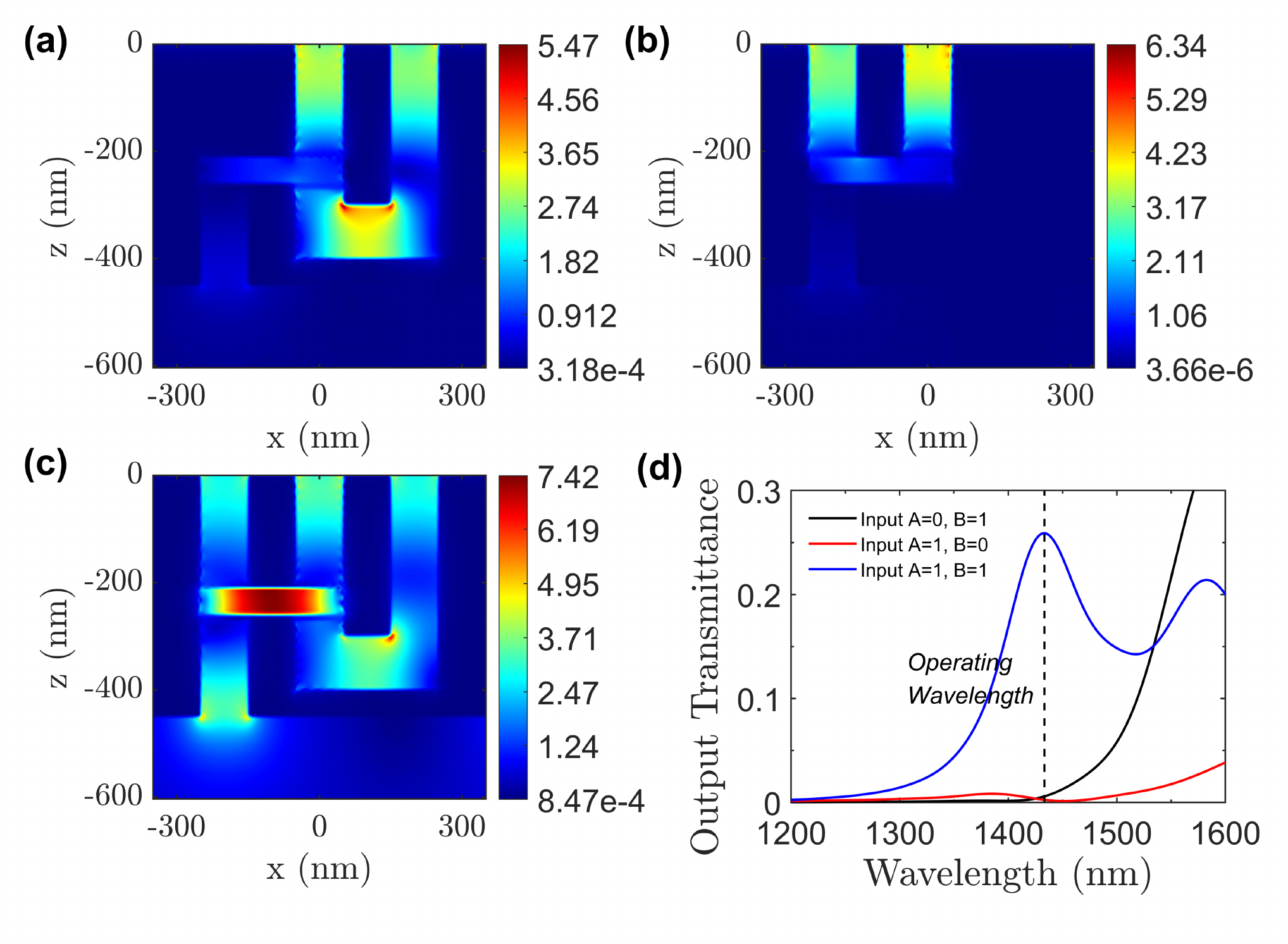}
    \caption{Electric field profile for AND gate representation: (a) Input port 1$->$ A=0, Input port 3 $->$ B=1, output=0 (b)  Input port 1 $->$ A=1, Input port 3 $->$ B=0, output=0 (c)  Input port 1 $->$ A=1, Input port 3 $->$ B=1, output=1 (d) Output transmission spectrum for all the inputs with maximum value of 27\% at 'ON' state and minimum of 1\% at 'OFF' state}
    \label{fig:8}
\end{figure}

From the transmittance curve in Fig.~\ref{fig:8}(d), it can be seen that the designed AND gate is operational at \(\lambda\) = 1440 nm. For output C = 1, 27\% transmittance is obtained at the output, whereas for C = 0, the transmittance is nearly 1\%. Contrast ratio for AND gate is measured 14.31 \(dB\). The electric Field profile from Fig.~\ref{fig:8}(a-c) validates the functionality of the AND gate.

\subsection{OR Gate}
For implementing the OR gate, port 1 is used for the first input(A), port 2 is used as the control port and port 3 is used for the second input(B). The output (C) is observed from port 4. Proper functionality is achieved only when the control pin is low (no light funnels inside port 2), which is listed in Table.~\ref{tab:3}.

\begin{figure}[h]
    \centering
    \includegraphics[width=1\linewidth]{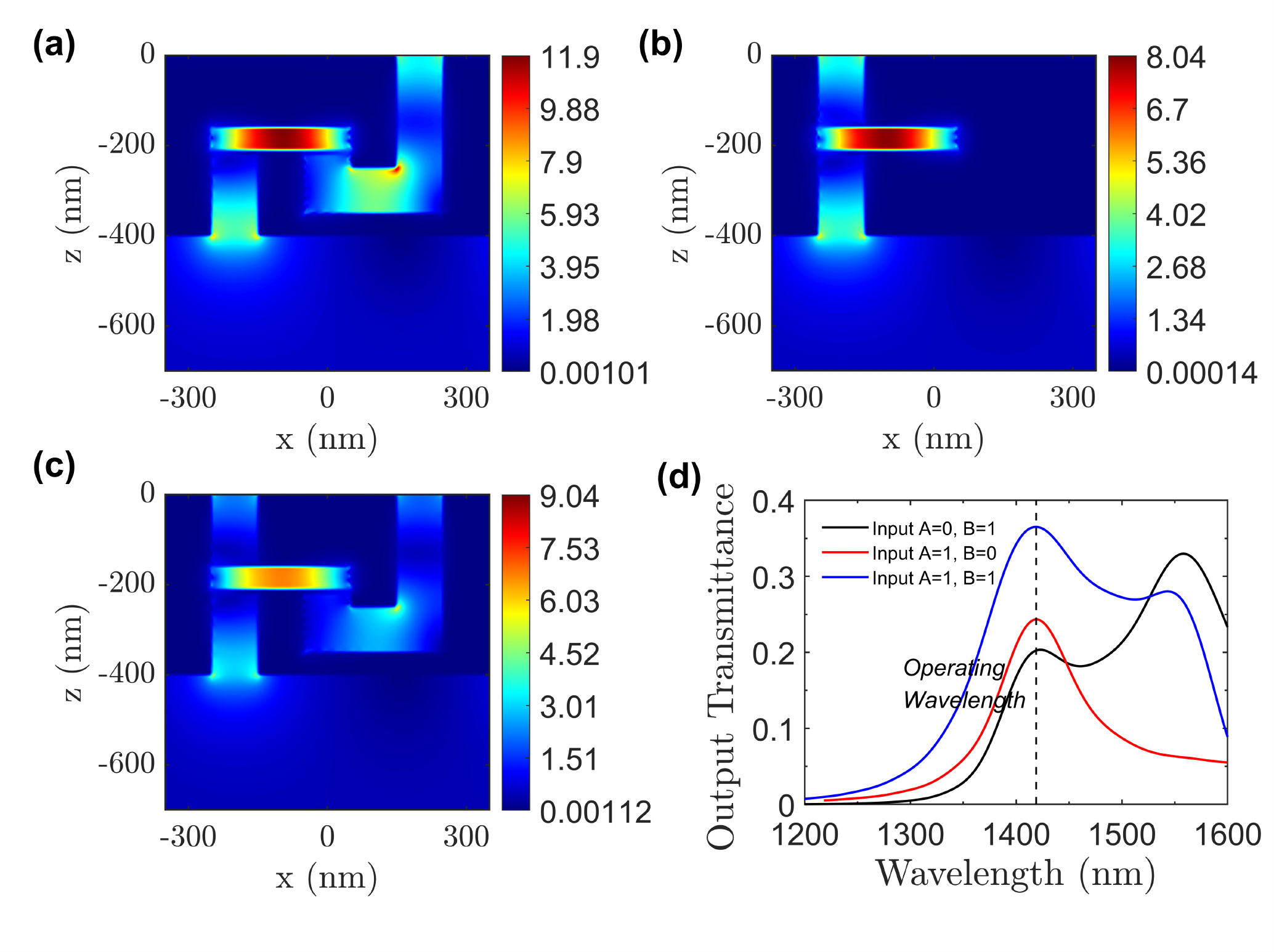}
    \caption{Electric field profile for AND gate representation: (a) Input port 1 $->$ A=0, Input port 3 $->$ B=1, output=1 (b)  Input port 1 $->$ A=1, Input port 3 $->$ B=0, output=1 (c)  Input port 1 $->$ A=1, Input port 3 $->$ B=1, output=1 (d) Output transmission spectrum for all the inputs with maximum value of 37\% at 'ON' state and minimum value at 'OFF' state is close to zero}
    \label{fig:9}
\end{figure}

From the transmittance curve in Fig.~\ref{fig:9}(d), it can be seen that the designed OR gate is operational at \(\lambda\) = 1418 nm. For output C = 1, maximum 37\% transmittance is obtained at the output when both inputs are ON, while for C = 0, the transmittance is nearly 0.5\%. The contrast ratio for the OR gate is found 18.69 \(dB\). The electric field profile from Fig.~\ref{fig:9}(b-d) validates the functionality of the OR gate.
\begin{table}[h!]
\centering
\caption{Truth table for implementing optical OR gate}
\renewcommand{\arraystretch}{1.5}
\begin{tabular}{|c|c|c|c|>{\centering\arraybackslash}p{1.7cm}|}
\hline
\shortstack{\rule{0pt}{3ex}\textbf{Control}\\\textbf{Port}} & \shortstack{\textbf{Input}\\\textbf{(A)}} & \shortstack{\textbf{Input}\\\textbf{(B)}} & \shortstack{\textbf{Output}\\\textbf{(C)}} &
\shortstack{\textbf{Contrast Ratio}\\\textbf{(T)}} \\
\hline
\multirow{4}{*}{Off} & 0 & 0 & 0 & \multirow{4}{*}{18.69 dB} \\
                      & 0 & 1 & 1 & \\
                      & 1 & 0 & 1 & \\
                      & 1 & 1 & 1 & \\
\hline
\end{tabular}
\label{tab:3}
\end{table}

Observing the transmittance spectra of the three gates NOT, AND and OR, it can be observed that the transmittance corresponding to output ’HIGH’ is relatively low because of the metallic absorption losses caused by \(Ag\) at the operating wavelength. Besides, bending inside the waveguide can cause scattering of the electromagnetic waves, resulting in energy loss and a significant amount of electric field is concentrated at the bending point. Therefore, considering the low transmission profiles at the gate outputs, a threshold transmittance of \(T_{th}\) = 20\% is set for distinguishing between \textbf{HIGH} and \textbf{LOW} output. Though the bending inside the waveguide causes energy loss, it helps to achieve a smaller footprint during circuit integration.

\section{Conclusion}
In this study, a compact and planar plasmonic metasurface-based structure was proposed and demonstrated to perform three fundamental all-optical logic operations—NOT, AND, and OR—within a single structure. Utilizing light funneling phenomenon in grooved MIM waveguides, the design achieves high contrast ratios up to 18.69 dB. By carefully designing the device parameters, the operation wavelength is confined to the 1400–1450 nm range, making it suitable for optical computation in telecommunications band. Its use of a single-directional light source for multiple input ports significantly reduces the footprint and enhances integration feasibility in photonic circuits. This work lays the foundation for more advanced, densely integrated, and energy-efficient optical logic systems, paving the way for ultrafast photonic processors.

\section*{Acknowledgments}
The authors acknowledge the financial grant provided by BUET (Basic Research Grant, Office Order no.: Shongstha/R-60/Re-2413, Dated: 10 Oct. 2023 (Professor Dr. Md. Zahurul Islam)) and logistical support provided by the Department of EEE, BUET throughout the duration of this work.

\newpage

\vfill


\begin{thebibliography}{1}
\bibliographystyle{IEEEtran}

\bibitem{ref1}
Y. A. Vlasov and S. J. McNab, “Losses in single-mode silicon-on-insulator strip waveguides and bends,” Opt. express 12, 1622–1631 (2004).

\bibitem{ref2}
S. Soma, S. K. C. Gowre, M. V. Sonth, et al., “Design and simulation of reconfigurable optical logic gates for integrated optical circuits,” Opt. quantum electronics 55, 340 (2023).

\bibitem{ref3}
A. M. Masoud, I. S. Ahmed, S. A. El-Naggar, and M. D. Asham, “Design and simulation of all-optical logic gates based on two-dimensional photonic crystals,” J. Opt. pp. 1–9 (2022).

\bibitem{ref4}
Y. Huang, M. Shi, A. Yu, and L. Xia, “Design of multifunctional all-optical logic gates based on photonic crystal waveguides,” Appl. optics 62, 774–781 (2023).

\bibitem{ref5}
F. Zhang, L. He, H. Zhang, et al., “Experimental realization of topologically-protected all-optical logic gates based on silicon photonic crystal slabs,” Laser \& Photonics Rev. 17, 2200329 (2023).

\bibitem{ref6}
L. He, D. Liu, H. Zhang, et al., “Topologically protected quantum logic gates with valley-hall photonic crystals,” Adv. Mater. 36, 2311611 (2024).

\bibitem{ref7}
T. He, H. Ma, Z. Wang, et al., “On-chip optoelectronic logic gates operating in the telecom band,” Nat. Photonics 18, 60–67 (2024).

\bibitem{ref8}
S. Zarei and A. Khavasi, “Realization of optical logic gates using on-chip diffractive optical neural networks,” Sci. Reports 12, 15747 (2022).

\bibitem{ref9}
I. S. Maksymov, “Optical switching and logic gates with hybrid plasmonic–photonic crystal nanobeam cavities,” Phys. Lett. A 375, 918–921 (2011).

\bibitem{ref10}
S. Kaur and R.-S. Kaler, “Ultrahigh speed reconfigurable logic operations based on single semiconductor optical amplifier,” J. Opt. Soc. Korea 16, 13–16 (2012).

\bibitem{ref11}
G.-Y. Oh, D. G. Kim, and Y.-W. Choi, “All-optical logic gate using waveguide-type spr with au/zno plasmon stack,” in OECC 2010 Technical Digest, (IEEE, 2010), pp. 374–375.

\bibitem{ref12}
Q. Xu and M. Lipson, “All-optical logic based on silicon micro-ring resonators,” Opt. express 15, 924–929 (2007).

\bibitem{ref13}
T. Liang, L. Nunes, M. Tsuchiya, et al., “High speed logic gate using two-photon absorption in silicon waveguides,” Opt. Commun. 265, 171–174 (2006).

\bibitem{ref14}
Z. Liu, L. Ding, J. Yi, et al., “Design of a multi-bits input optical logic device with high intensity contrast based on plasmonic waveguides structure,” Opt. Commun. 430, 112–118 (2019).

\bibitem{ref15}
Y. Dan, Z. Fan, X. Sun, et al., “All-type optical logic gates using plasmonic coding metamaterials and multi-objective optimization,” Opt. Express 30, 11633–11646 (2022).

\bibitem{ref16}
R. El Haffar, O. Mahboub, A. Farkhsi, and M. Figuigue, “All-optical logic gates using a plasmonic mim waveguide and elliptical ring resonator,” Plasmonics pp. 1–12 (2022).

\bibitem{ref17}
M. Moradi, M. Danaie, and A. A. Orouji, “All-optical nor and not logic gates based on ring resonator-based plasmonic nanostructures,” Optik 258, 168905 (2022).

\bibitem{ref18}
Y.-D. Wu, Y.-T. Hsueh, and T.-T. Shih, “Novel all-optical logic gates based on microring metal-insulator-metal plasmonic waveguides.” in PIERS proceedings, (2013).

\bibitem{ref19}
N. Nozhat and N. Granpayeh, “All-optical logic gates based on nonlinear plasmonic ring resonators,” Appl. optics 54, 7944–7948 (2015).

\bibitem{ref20}
K. Choudhary and S. Kumar, “Optimized plasmonic reversible logic gate for low loss communication,” Appl. Opt. 60, 4567–4572 (2021).

\bibitem{ref21}
F. Pardo, P. Bouchon, R. Haïdar, and J.-L. Pelouard, “Light funneling mechanism explained by magnetoelectric interference,” Phys. review letters 107, 093902 (2011).

\bibitem{ref22}
J. Le Perchec, P. Quémerais, A. Barbara, and T. López-Ríos, “Why metallic surfaces with grooves a few nanometers deep and wide may strongly absorb visible light,” Phys. Rev. Lett. 100, 066408 (2008).

\bibitem{ref23}
K. Ikeda, K. Takahashi, T. Masuda, et al., “Structural tuning of optical antenna properties for plasmonic enhancement of photocurrent generation on a molecular monolayer system,” The J. Phys. Chem. C 116, 20806–20811 (2012).

\bibitem{ref24}
E. D. Palik, Handbook of optical constants of solids, vol. 3 (Academic press, 1998).

\bibitem{ref25}
G. Xiao, Q. Zhu, Y. Shen, et al., “A tunable submicro-optofluidic polymer filter based on guided-mode resonance,” Nanoscale 7, 3429–3434 (2015).

\bibitem{ref26}
S. A. Maier et al., Plasmonics: fundamentals and applications, vol. 1 (Springer, 2007).

\bibitem{ref27}
F. Hu, H. Yi, and Z. Zhou, “Band-pass plasmonic slot filter with band selection and spectrally splitting capabilities,” Opt. express 19, 4848–4855 (2011).

\bibitem{ref28}
M. J. Madou, Fundamentals of microfabrication: the science of miniaturization (CRC press, 2018).

\bibitem{ref29}
S. Roy, Y. Gupte, and T. Green, “Flow cell design for metal deposition at recessed circular electrodes and wafers,” Chem. engineering science 56, 5025–5035 (2001).

\bibitem{ref30}
M. Köhler, Etching in microsystem technology (John Wiley \& Sons, 2008).

\bibitem{ref31}
T. Green, “Gold etching for microfabrication,” Gold bulletin 47, 205–216 (2014).

\bibitem{ref32}
M. A. Lieberman and A. J. Lichtenberg, “Principles of plasma discharges and materials processing,” MRS Bull. 30, 899–901 (1994).

\bibitem{ref33}
F. F. Chen and J. P. Chang, Lecture notes on principles of plasma processing (Springer Science \& Business Media, 2003).

\bibitem{ref34}
S. Kim, Y. Xuan, V. P. Drachev, et al., “Nanoimprinted plasmonic nanocavity arrays,” Opt. Express 21, 15081–15089 (2013).

\bibitem{ref35}
Z. Yuan, N. H. Dryden, J. J. Vittal, and R. J. Puddephatt, “Chemical vapor deposition of silver,” Chem. materials 7, 1696–1702 (1995).

\bibitem{ref36}
M. Kariniemi, J. Niinisto, T. Hatanpaa, et al., “Plasma-enhanced atomic layer deposition of silver thin films,” Chem. Mater. 23, 2901–2907 (2011).

\bibitem{ref37}
M. Makela, T. Hatanpaa, K. Mizohata, et al., “Studies on thermal atomic layer deposition of silver thin films,” Chem. Mater. 29, 2040–2045 (2017).

\bibitem{ref38}
MicroChemicals GmbH, “Application notes,” https://www.microchemicals.com/DOWNLOADS/Application-Notes/ (2025). Accessed: 2025-07-15.

\bibitem{ref39}
Y. Lee, S. Seo, A. Shearer, et al., “Hfo2 area-selective atomic layer deposition with a carbon-free inhibition layer,” Chem. Mater. 36, 4303–4314 (2024).

\bibitem{ref40}
Y. Choi, T. Kim, H. Lee, et al., “Bottom-up plasma-enhanced atomic layer deposition of sio2 by utilizing growth inhibition using nh3 plasma pre-treatment for seamless gap-fill process,” Sci. Reports 12, 15756 (2022).


\end{thebibliography}
\end{document}